\documentclass[conference]{IEEEtran}
\IEEEoverridecommandlockouts
\usepackage{cite}
\usepackage{amsmath,amssymb,amsfonts}
\usepackage{algorithmic}
\usepackage{graphicx}
\usepackage{textcomp}
\usepackage{xcolor}
\usepackage{blindtext, graphicx}
\usepackage{listings}
\usepackage{graphicx}
\usepackage[font=small]{caption}

\usepackage[utf8]{inputenc}

\def\BibTeX{{\rm B\kern-.05em{\sc i\kern-.025em b}\kern-.08em
    T\kern-.1667em\lower.7ex\hbox{E}\kern-.125emX}}

\begin{document}

\title{ Development of An  Assessment Benchmark for Synchronous Online Learning for Nigerian Universities}



\author{
\IEEEauthorblockN{Modesta Ezema\textsuperscript{1}, Boniface Nworgu\textsuperscript{2,3}, Deborah Ebem\textsuperscript{1}, Stephenson Echezona\textsuperscript{1}, \\ Celestine Ugwu\textsuperscript{1}, Assumpta Ezugwu\textsuperscript{1}, Asogwa Chika\textsuperscript{1}, Ekene Ozioko\textsuperscript{4}, Elochukwu Ukwandu\textsuperscript{5}}.

\IEEEauthorblockA{\textsuperscript{1}Dept. of Computer Science,  Faculty of Physical Science, University of Nigeria, Nsukka, Nigeria.\\\textsuperscript{2}Dept. of Science Education, Faculty of Education, University of Nigeria, Nsukka, Nigeria.\\\textsuperscript{3}Centre for Distance and E-Learning, University of Nigeria, Nsukka, Nigeria.\\\textsuperscript{4}School of Mathematical, Physics and Computational Science, University of Reading, United Kingdom.\\\textsuperscript{5} Dept. of Computer Science, Cardiff School of Technologies, Cardiff Metropolitan University, United Kingdom.\\
}

E-mail addresses: {(modesta.ezema, boniface.nworgu, debora.ebem, stephenson.echezona, celestine.ugwu}\\
{assumpta.ezugwu, chika.asogwa.181578)}@unn.edu.ng \\ e.f.ozioko@pgr.reading.ac.uk, eaukwandu@cardiffmet.ac.uk \\
Corresponding author: Elochukwu Ukwandu
}

\maketitle
\begin{abstract}

In recent times, as a result of COVID-19 pandemic, higher institutions in Nigeria have been shutdown and the leadership of Academic Staff Union of University (ASUU) said that Nigerian universities cannot afford to mount Online learning platforms let alone conduct such learning system in Nigeria due to lack of infrastructure, capacity and skill sets in the face of COVID-19 pandemic. In the light of this, this  research undertook an online survey  using University of Nigeria, Nsukka (UNN) as a case study  to know which type of online learning system ASUU leadership is talking about - Asynchronous or Synchronous? How did ASUU come about their facts? Did ASUU base their assertion on facts, if YES, what are the benchmarks? Therefore, this research  project is focused on providing benchmarks to assess if a Nigerian University has what it takes to run a synchronous  Online Learning. It includes Infrastructure needed (Hardware, Software, Network connectivity), Skill sets from staff  (Computer literacy level). In a bid to do this, an online survey was administered to the staff of Centre for Distance and E-learning of UNN and out of the 40  members of that section of the University, we had 32 respondents. The survey seeks to find whether UNN has the requisite infrastructure  and the skill sets to mount synchronous online learning. The results of the study reveal that the infrastructure domain of the questionnaire consist of 13 questions. Therefore, the maximum  scores that a respondent can have is 13. From Table~\ref{tab: table2} below, only 3 (9.4\%) respondents scored 13. This shows that the rating for excellence is quite low among the population sampled, hence we conclude that  University of Nigeria, Nsukka does not have the infrastructure  to mount synchronous online learning.  The Skills sets domain of the questionnaire consist of 18 questions. Therefore, the maximum  scores that a respondent can have is 18. From Table~\ref{tab: table1} below, only 8 (25\%) respondents scored 18. This shows that the rating for excellence is quite low among the population sampled as they were not even up to 50\%. This shows that University of Nigeria, Nsukka does not have the skill sets required to mount synchronous online learning.

\textbf{Keywords}: online learning, infrastructure, skill set, university, benchmark 

\end{abstract}

 \section{\textbf{Introduction}}
 \label{Section: Introduction}
   Synchronous online learning is one aspect of virtual learning, which involves using computer technologies to provide real-time virtualised education to those who due to various reasons are unable to enroll for full face-to-face academic programmes in  the University. Given computer access, it is believed that, people can then use the Information and Communication Technology (ICT) tool to do all sorts of things that improve their socio-economic status. This form of education has undeniably made learning more flexible and convenient in terms of time, cost and what is on offer. However, in spite of the obvious advantages of this form of education, the issue of standard and quality control remains paramount.\\

   According to Ilechukwu~\cite{ilechukwu2014optimizing}, the issue of quality assurance in online learning was not in place until 2009 when the National Universities Commission (NUC) in charge of the quality assurance in Nigerian universities established a unit to ensure quality in online learning according to Owoye~\cite{owoeye2009quality}. The spontaneous growth of computers, the internet and other electronic devices provide global opportunities for education, especially for learning outside the school premises, according to~\cite{mmeremikwu2018assessment}. \\
   A key aspect of the online learning is that it does not rank institutions, but rather acknowledges the reality that all institutions will have aspects of strength and weakness that can be learnt from and improved. The rapid growth in the technologies being used, the ways that they are being applied across an ever widening group of academic disciplines and the evolving skills and experience of teachers and students means that online learning is a moving target. Any benchmarking approach that presumes particular online learning technologies or pedagogies is unlikely to meaningfully assess a range of institutions within a single country and beyond, according to marshal~\cite{marshall2007benchmarking}.\\ This research through questionaire analysis is trying to find out whether Universities in Nigeria has what it takes to do synchronous online learning using University of Nigeria, Nsukka as case study.  \\
 
 \section{\textbf{Literature Review}}
 \label{Section: Literature Review}
 \subsection{\textbf{The concept of online learning}}
Eze~\textit{et al.} in~\cite{eze2018utilisation} defined online learning as an aggregate of a digitally empowered learning framework that utilises hardware for example, Personal Computers, Tablets, digital cameras, videos, overhead projector and so on. Software, such as: operating systems, cloud technologies, applications, etc and others such as: Universal Serial Bus (USB) drives, Compact Disk textbooks, electronic content, etc., either from a distance or face-to-face classroom setting to empower and facilitate teacher to student interactions. In essence, online learning moves the citadel of learning from the traditional approach - (teacher-focused), to a module-driven, Information and Communication Technology (ICT)-based customisable learning approach (student-focused). Online learning provides the framework and empowers students and teachers to produce, progress, and share learning content in a more regular structure \cite{wallace2003online}.\\

 Raja~\textit{et al.} in~\cite{solaja2019impact} lists the advantages of ICT and online learning in education as follows:
\begin{itemize}
 \item It makes the learning process more exciting for the students.
\item Helps students learn at their own pace, especially for those in work and study programmes. 
\item It can serve as a great skill acquisition process for students to learn new ICT skills they can use later in the workplace.  
\item Decrease paper and photocopying costs, which is also good for the environment and promotes the "green revolution" concept.\\
\end{itemize}

\subsection{\textbf{Online learning adoption in Nigerian institutions}} 
Each year, a very high percentage of prospective students seeking admission into Nigerian higher education institutions are turned back. At the same time, those admitted are congested in large classrooms benefiting little or nothing from their lectures. Online learning has emerged as a solution to these challenges. Research indicates that computer technology can help support learning and useful in developing the higher-order skills of critical thinking, analysis, and scientific inquiry~\cite{jeremy2000changing}.

The authors in~\cite{anene2014problem} investigated the availability of facilities for e-learning and e-learning materials in Nigerian Universities. From their study, a total of two hundred and twenty-eight (228) students constituted the sample for the study in the six geopolitical zones of Nigeria. They used percentages statistics in analysing data of the study. Their study shows that in Nigeria, a formidable obstacle to using ICT in education and other sectors is infrastructure deficit, known as digital poverty. The majority of the students reported that their Universities do not have an e-learning library domain, neither do they organise online seminars (webinars), discussions and examination. These are perhaps due to limited bandwidth within their school environment as well as the requiste technical know-hows. The authors in~\cite{eze2018utilisation} also listed the instructors' inability to assist students in building up capacity and information needed to adequately utilise e-learning facilities as a significant barrier for online learning adoption in Nigeria.\\

\subsection{\textbf{Review of related literature}}
Different researches have been published in this field. In this section, we will present a few of this literature.
Wilson in~\cite{nwankwo2018promoting}, reviewed the current state of affairs in Nigeria's University education policies. The author emphasised the urgent need to revitalise the deteriorating system by introducing and standardisation of online learning platforms supported by cutting-edge information technology tools. The author investigated some vital issues with online learning in Nigeria, including:

\begin{itemize}
\item	Whether or not the Nigerian government has laws and policies to support equitable access to education at the university level? 
\item	Whether or not the purpose of University education in Nigeria could be satisfied through online learning systems? 
\item	Whether or not an online learning system could be rightly run to offer an alternative mode of learning and complement the traditional university system. These are in relation to the technological advancements around the globe and the promotion and acceptance of such mode of learning in developed countries?
\end{itemize}

In~\cite{okorafor2012developing}, the authors developed an e-learning strategy for open and distance education programmes in Nigeria. They argued that the problems confronting Nigerian education, especially the higher education sector, demand a rethinking in the design of learning experiences, courses and delivery mode, teacher-student contact, and the academic role. Their e-learning framework was tagged AASCE Framework for Quality E-learning Strategy, while the AASCE is an acronym for Administration, Analysis, Institutional Strategy, Courseware Development, and Evaluation.\\

The author in~\cite{al2016empirical} investigated e-learning acceptance and assimilation of students in the King Khalid University, Saudi Arabia. The author surveyed 286 participants (students) and used structural equation modeling for data analysis to determine the factors influencing the learners' intention to use e-learning. The result shows the predicting (promoting/inhibiting) factors of e-learning technology acceptance, while also examining some related post-implementation interventions expected to contribute to the acceptance and assimilation of e-learning systems. The results also outline valuable outcomes such as managerial interventions and controls for better organisational e-learning management, leading to greater acceptance and effective utilisation.\\

In~\cite{evoh2007policy}, the author examined the prospects and challenges of the New Partnership for Africa's Development (NEPAD) e-schools initiative. The author evaluated the prospects and challenges the e-schools face in Africa's collaborative initiatives under the auspices of the NEPAD. The author identified the need for political goodwill and African governments' support for the project to be fully realised. \\

Liverpool~\textit{et al.} in~\cite{liverpool2009towards} detailed the experience the University of Jos faced in its six-year effort to implement an e-learning initiative in Mathematics with support from the Carnegie Corporation of New York, Hewlett Packard, and the World Bank-sponsored Science and Technology Education Post-Basic (STEPB) Project. \\

In~\cite{price2007framework}, the authors developed a model framework for studying technology's impact on education. Their research's primary outcome was to demonstrate the usefulness of the three-part model for evaluating the impact of technology in education. \\

\textbf{The three-part model consists of:}
\begin{itemize}
 \item	Anticipatory (such as the discourses and rhetoric of policy, design and intentions, opinion and attitudes); 
 \item	Ongoing (processes of integration, including practices of staff development); and 
 \item	Achieved (summative studies, particularly of technology that is no longer considered 'novel')
\end{itemize}

Alabi in~\cite{olanike2010education}, investigated the Nigerian undergraduate students' readiness for e-learning. The authors collected data with the use of questions and observation methods from 1,450 undergraduate students of the University of Lagos, Nigeria. The study showed that students are ready for e-learning, and have the right learning attitude. The study also revealed that about half of the students have laptop as one of the personal technologies for e learning.\\

Shonola~\textit{et al.} in~\cite{shonola2016impact}, examined the impact of the educational use of mobile devices/technology in supporting the learning process in Nigerian Universities. They sampled 240 university students with questionnaires with the aim of exploring students' interactions with their portable gadgets. Their result showed increased use of mobile technology for academic-related activities by the students.\\

Agbo in~\cite{agbo2015factors} investigated the factors that hinder the use of ICT in teaching and learning Computer studies in Ohaukwu Local Government Area of Ebonyi State. The study was carried out on ten (10) secondary schools in Ohaukwu Local Government Area and indicated some influential factors like; the low level of accessibility and cost of ICT equipment, lack of ICT training to teachers, Teachers and Students attitude towards computer usage, and lack of parents and community support as the major factors hindering the use of ICT in teaching and learning of computer studies in Ohaukwu Local Government Area of Ebonyi State.\\

Mojaye~\cite{mojaye2015mobile} examined mobile phones' evolution in Nigeria and its positive and negative impact on Nigerian higher education institutions' students. The author listed some of the positives of mobile phone usage by students to include; easy access to information and mobile phones as a convenient teaching tool. The author also listed classroom distraction, cheating, and reduced cognitive ability as some of the Adverse effects of mobile phone usage among students.\\

Adedokun-Shittu in~\cite{adedokun2015assessing} examined the impact of ICT deployment in teaching and learning in Nigeria. In the study, survey data was sampled from 593 respondents (students and lecturers) using a mixed-method design consisting of qualitative and quantitative procedures. This study produced a model conceived as a conceptual framework for researchers on impact assessment and comprising; Positive effect, Integration, Incentives, and Challenges.\\

From the above analysis none has thought of the  developing  assessment benchmark for online learning in Nigerian universities.

\section{\textbf{Materials Selection}}
\label{Section: Material Selection}
The material selection for this work involved an ordered and systematic process in analysing and eliminating unsuitable materials and identifying the ones which are the most suitable for this analysis. The material used for this analysis involved the design, development, distribution, and response collection of questioners containing the parameters of e-learning indicators. 

\section{\textbf{Research Method}}
\label{Section: Research Method}
The goal of this assessment benchmark is to formulate strengths, weaknesses, and areas for enhancement through an analysis of the current status to suggest desirable states of affairs and contributes to the e-learning transformation process.
The questionnaire design method suggested by~\cite{zaharias2009developing} was used. 
 The strategies utilised in the collection of data for this analysis to uncover the assessment benchmark for synchronous online learning involve 32 sample questionnaires. This methodology contains 3 stages: 
 
\begin{itemize}
\item  Development of the questionnaire: The analysis parameters used in developing the questions were measured with the approved Nigerian University Commission's benchmark adapted for e-learning.
\item Question distribution and collection from respondents: The sample size used were 32 staff of Center for distance learning University of Nigeria Nsukka.
\item Comparative analysis of the results: The response from the questionnaire were analysed using the Nation Universities Commission (NUC) benchmark to inform our recommendation.   
\end{itemize}

  \section{\textbf{Results}}
  \label{Section: Results}
 This section presents the result of our analysis. The following hypotheses were tested: 
 
 H0: University of Nigeria, Nsukka has the skill set to mount synchronous online learning.
 
 H1: University of Nigeria, Nsukka does not have the skill set to mount synchronous online learning.
 
 H20: University of Nigeria, Nsukka has the infrastructure to mount synchronous online learning. 
 
 H21: University of Nigeria, Nsukka does not have the infrastructure to mount synchronous online learning. Table 1 below the result of the skill set readiness of University of Nigeria, Nsukka.   
 
 \begin {table}[h!]
  \begin {center}
   \caption{ Skill Sets Assessment Scores for Synchronous online learning}
      \label{tab: table1}
       \begin{tabular}{l|c|c|c|c}
         \textbf{Scores} & \textbf{Frequency} & \textbf{Percent}
& \textbf{Valid Percent } & \textbf{Cumulative Percent}\\
         \hline
          3 & 2 & 6.3 & 6.3 & 6.3\\
          8 & 1 & 3.1 & 3.1 & 9.4\\
          9 & 1 & 3.1 & 3.1 & 12.5\\
         10	& 2 & 6.3 & 6.3	& 18.8\\
         12 & 2 & 6.3 & 6.3 & 25.0\\
         13 & 2 & 6.3 & 6.3 & 31.3\\
         14 & 2 & 6.3 & 6.3 & 37.5\\
         15 & 4 & 12.5 & 12.5 & 50.0\\
         16 & 7 & 21.9 & 21.9 & 71.9\\
         17 & 1 & 3.1 & 3.1 & 75.0\\
         18 & 8 & 25.0 & 25.0 & 100.0\\
         Total & 32 & 100.0 & 100.0\\
      \end{tabular}
  \end {center}
\end {table}		

The skills sets domain of the questionnaire consist of 18 questions. Therefore, the maximum score that a respondent can have is 18. From the Table \ref{tab: table1} above, only 8 (25\%) respondents scored 18. This shows that the rating for excellence is quite low among the population sampled. They were not even up to 50\%. This shows that University of Nigeria, Nsukka does not have the skill set to mount synchronous online learning. Table \ref{tab: table2} below gives the result for infrastructural  readiness.  

\begin {table}[h!]
  \begin {center}
   \caption{ Infrastructure Assessment scores for Synchronous online learning}
      \label{tab: table2}
       \begin{tabular}{l|c|c|c|c}
         \textbf{Scores} & \textbf{Frequency} & \textbf{Percent}
& \textbf{Valid Percent} & \textbf{Cumulative Percent}\\
         \hline
3 & 1 & 3.1 & 3.1 & 3.1\\
4 & 3 & 9.4 & 9.4 & 12.5\\
5 & 2 & 6.3 & 6.3 & 18.8\\
6 & 1 & 3.1 & 3.1 & 21.9\\
7 & 3 & 9.4 & 9.4 & 31.3\\
8 & 4 & 12.5 & 12.5 & 43.8\\
9 & 3 & 9.4 & 9.4 & 53.1\\
10 & 1 & 3.1 & 3.1 & 56.3\\
11& 7 & 21.9 & 21.9 & 78.1\\
12 & 4 & 12.5 & 12.5 & 90.6\\
13 & 3 & 9.4 & 9.4 & 100.0\\
Total & 32 & 100.0 & 100.0\\
  \end{tabular}
 \end {center}
\end {table}

The infrastructure domain of the questionnaire consist of 13 questions. Therefore, the maximum scores that a respondent can have is 13. From Table \ref{tab: table2} above, only 3 (9.4\%) respondents scored 13. This shows that the rating for excellence is quite low among the population sampled. They were not even up to 25\%. This shows that University of Nigeria, Nsukka does not have the infrastructure to mount online learning. 

 \section{\textbf{Result Discussions}}
 \label{Section: Result Discussion}
  This study tends to assess the infrastructural and skills readiness of Nigerian universities to mount online learning using University of Nigeria, Nsukka as case study. Infrastructure is one of the benchmark for the actualisation of online learning in any university worldwide. Nine items were considered under infrastructure to assess the readiness of Nigerian universities to mount online learning. We present the descriptive findings of participants’ assessment. The responses of the 32 respondents under the infrastructure category are presented in Table \ref{tab: table1}. The result shows that 16 respondents which represent 50\% of all the respondents agreed that during COVID-19, the entire learner support infrastructure at University of Nigeria, Nsukka had to change from mostly face-to-face to completely online learning while 15 respondents which represent 46.9\% of all the respondents disagreed and 1 respondent which represent 3.1\% of all the respondents were indifferent. \\
  
The result shows that 22 respondents, which represent 68.8\% of all the respondents agreed that University of Nigeria, Nsukka has e-learning laboratory for courseware development, while only 4 respondents which represent 12.5\% of all the respondents disagreed and 6 respondents which represent 18.8\% of all the respondents were indifferent. Also, 20 respondents which represent 62.5\% of all the respondents agreed to have made use of the following authoring tools during the courseware development: MS Powerpoint, Microsoft Word processor, text, audio, animation, others while 10 respondents which represent 28.1\% of all the respondents disagreed and 2 respondents which represent 6.3\% of all the respondents were indifferent. Availability of steady internet is vital to online learning and the result indicates that University of Nigeria, Nsukka can afford steady supply of broadband to drive online learning. 23 respondents, which represent 71.9\% of all the respondents agree that University of Nigeria, Nsukka can afford steady supply of broadband to drive online learning, while 7 respondents which represent 21.9\% of all the respondents disagree and 2 respondents which represent 6.3\% of all the respondents were indifferent. On the availability of adequate internet equipment, 16 respondents which represent 50.1\% of all the respondents agreed that University of Nigeria, Nsukka possesses enough Internet equipment to drive e-learning while 14 respondents which represent 43.8\% of all the respondents disagreed and 2 respondents which represent 6.3\% of all the respondents were indifferent. \\

On the issue of power supply, 19 respondents which represent 59.4\% of all the respondents agreed that there is poor power supp in University of Nigeria, Nsukka while 11 respondents which represent 34.4\% of all the respondents disagreed and 2 respondents which represent 6.3\% of all the respondents were indifferent. This suggests that there is poor power supply in University of Nigeria, Nsukka. Computer based test center is a key infrastructure to be considered in assessing the readiness of a university to mount online learning especially for examination purpose. The result of this study reveals that 30 respondents which represent 93.7\% agree that University of Nigeria, Nsukka have Computer based Test Center while 1 respondent which represent 3.1\% of all the respondents disagree and 2 respondents which represent 6.3\% of all the respondents were indifferent. This result suggests that University of Nigeria Nsukka have Computer based Test Center. \\

This study further reveals the capacity of the Computer-based Test Center. 25 respondents which represent 78.1\% of all the respondents agree that more than one department can conduct examination concurrently in the computer based center while 6 respondent which represent 15.6\% of all the respondents disagree and 1 respondent which represent 3.1\% of all the respondents were indifferent This suggests that the Computer-based Test Center can accommodate several departments concurrently during any examination.\\

For the skill set, 2 items were considered to assess the readiness of Nigerian universities to mount online learning. The responses of the 32 respondents under the skill set category are presented in Table \ref{tab: table2}. The result shows that 27 respondents, which represent 84.4\% of all the respondents agreed that University of Nigeria, Nsukka can boast of qualified Technologists and Engineers to man the ICT in their various campuses while 3 respondents which represent 9.4\% of all the respondents disagreed and 2 respondents which represent 6.3\% of all the respondents were indifferent. Also, 27 respondents which represent 84.4\% of all the respondents agreed that University of Nigeria, Nsukka can boast of qualified content editors across all fields to drive the courseware for e-learning while 3 respondents which represent 9.4\% of all the respondents disagreed and 2 respondents which represent 6.3\% of all the respondents were indifferent. \\

Moreover, in order to determine the readiness of the University for Synchronous Online Learning, both infrastructure and skill set assessment scores were used, and the benchmark for the score for each of the component is 100\%. \\

From the result, it was discovered that only 8 respondents out of 32 as shown in Table \ref{tab: table1} scored the University 100\% in the aspect of skill set. This number just represents only 25\% of the entire respondents. Hence, University of Nigeria, Nsukka is not ready in the aspect of skill set for Synchronous Online Learning.\\

It was equally found from the result that only 3 respondents as shown in Table \ref{tab: table2} scored the University 100\% in the area of infrastructure. The 3 respondents here represent only 9.4\% of the sample size which shows also that University of Nigeria, Nsukka is still not ready for Synchronous Online Learning in terms of infrastructure.\\

Since the percentage representation of the respondents that scored the University 100\% is very far below the 100\% benchmark in terms of skill set and infrastructure, we conclude that University of Nigeria, Nsukka is not ready for Synchronous Online Learning.\\

 \section{\textbf{Recommendation and future work}}
 \label{Section: Recommendation and Future Work}
 With the outbreak of coronavirus disease 2019 (COVID-19) and the fast spread of the disease across the globe, there is an urgent need for the stakeholders in education sectors to redefine the teaching and learning methods, more especially in developing countries like Nigeria. The government, education board, teachers and students need to come together to strategise on how to shift from the traditional face-to-face learning to online learning or blended learning. This will help to curtail the spread of the disease and also forestall the closing down of schools occasioned by the virus. To actualise this dream, adequate provision of online learning facilities and continuous training and retraining lecturers, students and all other personnel involved in teaching and learning must be ensured.\\ 
 
 In this paper, a pilot study was conducted to find out how ready universities in Nigeria are for online learning using University of Nigeria, Nsukka as a case study. The direction of our future study will be to carry out the updated version of the survey in which other universities with online learning approval from Nigerian Universities Commission (NUC) will be incorporated into our research scope. Moreover, an experimental research to establish the readiness of these universities or their level of preparedness as regards online learning will also be embarked upon in future work. 
 
 \section{\textbf{Conclusion}}
 \label{Section: Conclusion}
 In recent times, as a result of COVID-19 pandemic, higher Institutions in Nigeria were shut down and the leadership Academic Staff Union of Nigerian Universities (ASUU) claimed that Nigerian University cannot afford to mount online learning platform because of lack of infrastructure, capacity and skill set. The research provided benchmark to assess whether Nigerian Universities have what it takes to run an online learning. The study used University of Nigeria, Nsukka as a case study. From the result gotten from the questionnaire analysis, University of Nigeria, Nsukka do not have the skill set to mount online learning because from the results shown in Table~\ref{tab: table1} above, only 8(25\%) of respondents scored 18. This shows that the rating for excellence is quite low among the sampled population.  \\
 Also on the infrastructure level, only 3(9.4\%) respondents scored 13. This shows that the rating for excellence is quite low among the population sampled. From these results, University of Nigeria, Nsukka does not have staff member with the requisite minimum skill set nor infrastructure to mount online learning.\\

 \bibliographystyle{IEEEtran}
 \bibliography{Bibliography.bib}
 
 \end{document}